\documentclass[preprint,5p,twocolumn]{elsarticle}
\usepackage{lineno,hyperref}
\usepackage{booktabs,makecell,multirow}
\usepackage{lineno,hyperref}
\usepackage{amsmath}
\usepackage{amsthm}
\usepackage{amssymb}
\usepackage{indentfirst}
\usepackage{makecell}
\usepackage{enumerate}
\usepackage{array}
\usepackage{caption} 
\usepackage{algorithm}
\usepackage[noend]{algorithmic}

\usepackage{float}  
\usepackage{lipsum}
\usepackage{fancyhdr}
\usepackage[caption=false,font=footnotesize]{subfig}


\begin{document}

\begin{frontmatter}

\title{Efficient and Fine-grained Redactable Blockchain Supporting Accountability and Updating Policies}

\author[1]{Bin Luo\corref{cor1}}
\ead{xsbinluo@163.com}

\cortext[cor1]{Corresponding author}

\address[1]{College of Information Science and Technology, Jinan University, Gangzhou, China, 510632}

\begin{abstract}
    Redactable Blockchain aims to ensure immutability of the data for most of appications, and provide authorized mutability for some specific applications such as removing illegal content from blockchains. However, the existing redactable blockchain scheme has low redacting efficiency, and lacks an accountable and updatable fine-grained mechanism to control redacting rights.
    To solve the above problems, we propose an efficient and fine-grained redactable blockchain scheme with accountability and updatable policies, called \emph{EFRB}. In our scheme, the transaction owner can set the updatable policy associated with the transaction, and only users who meet the policy can become the redactor of the transaction.
    In order to prevent redactors from abusing redacting right, we introduce the concept of a witness group. A redacted transaction is legal if and only if it contains the signatures of redactor and witness group. We first give the concept of witness group, and then show the proposed \emph{EFRB} scheme. Finally, we demonstrate that \emph{EFRB} scheme is feasible and efficient through a series of experiments and analysis.
\end{abstract}

\begin{keyword}
\texttt Redactable blockchain\sep Chameleon hash\sep Digital signature\sep Accountability
\end{keyword}

\end{frontmatter}

% \linenumbers

\section{Introduction}

Blockchain can be regarded as a decentralized database and a distributed ledger technology. Compared with traditional databases, blockchain has the important properties of decentralization and immutability. Decentralization effectively solves the problem of single point of failure, and immutability effectively ensures the integrity of data.
But everything has two sides. On the one hand, these superior properties enable the blockchain to build a trust bridge for nodes in an untrusted environment and ensure the authenticity and reliability of data. Therefore, industry and commerce consider using blockchain technology to solve many practical problems, such as healthcare \cite{tanwar2020blockchain}, logistics \cite{hackius2017blockchain}, e-government \cite{hou2017application}.
On the other hand, in the process of application deployment, more and more cases show that immutability limits the development of blockchain. For example, institutions that use blockchain technology to provide services need to manage user data and modify it when necessary, but the immutability of the blockchain does not allow redacting these data. Secondly, if vulnerabilities are found in the smart contract after it is created, the existing methods can only repair the vulnerabilities through hard fork \cite{tsankov2018securify},
but this will affect the relevant transaction results and destroy the stability of the blockchain system. Thirdly, many reports mentioned that there are some illegal data in the blockchain, especially in bitcoin \cite{matzutt2018quantitative}. Because the blockchain cannot be redacted, it becomes very difficult to change wrong data or delete illegal data.
Finally, the EU's General Data Protection Regulation (GDPR) and other regulations require citizens to be given the "Right to be Forgotten", which cannot be met by the current blockchain \cite{schwerin2018blockchain}. Therefore, in order to solve the problems caused by the immutability of the blockchain in the application process, it is necessary to design an efficient and fine-grained method to allow data in the blockchain to be redacted in special scenarios.

\subsection{Related Work}
In recent years, scholars began to pay attention to these problems caused by the immutability of blockchain, gradually studied the redactable function of blockchain. In the currently proposed scheme of redactable blockchain, it is mainly divided into two redacting consensus methods.

One is based on voting, which requires the number of votes to reach a certain threshold to reach redacting consensus, and its core is how to choose legal voters. In 2019, Deuber et al.\cite{deuber2019redactable} proposed a redactable blockchain scheme in the permissionless setting based on the voting mechanism, which stipulated that miners were voters, and adds an "old state" in the block to maintain the integrity of the hash chain. However, it took too long to reach a redacting consensus in this scheme. In order to solve this problem,
Li et al.\cite{li2021escaping} proposed a instantly redactable scheme for Proof-of-Work (PoW) and Proof-of-Stake (PoS) blockchain. In this scheme, a virtual difficult problem is used to select the committee with voting rights. However, this scheme and the previous scheme are only coarse-grained redacts, and there is no corresponding voting incentive mechanism. In addition, Thyagarajan et al.\cite{thyagarajan2021reparo} proposed a redactable protocol Reparo,
which can repair vulnerable smart contracts and remove illegal content from the chain. But, the protocol may need to cascade modify the affected subsequent transactions when necessary.

The second is based on chameleon hash function, which is a one-way hash function with trapdoor proposed by Krawczyk and Rabin \cite{krawczyk1998chameleon}, and its core is how to manage the trapdoor. This function allows users with trapdoors to find a pair of hash collisions, that is, different function inputs can correspond to the same output value. The first redactable scheme was proposed by Ateniese et al.\cite{ateniese2017redactable} in 2017.
The scheme uses chameleon hash function instead of the original hash function of blockchain, which can modify blockchain data while keeping the hash value unchanged. However, this is a block level redacting scheme, which means that even changing a transaction requires the participation of the entire block data. To solve this problem, Derler et al.\cite{derler2019fine} introduced the concept of policy-based chameleon-hashes (PCHs) and proposed a fine-grained redactable scheme.
In this scheme, the transaction owner can set the policy of redacting the transaction, encrypt the trapdoor of the chameleon hash function and embed it in the policy. Only users whose attributes meet the policy can decrypt and obtain the trapdoor to redact the transaction. However, once the user obtains the trapdoor, he can redact data many times, and there is no corresponding trapdoor revocation mechanism. Later, Tian et al.\cite{tian2021accountable} proposed a fine-grained rewriting scheme in the permissionless blockchain.
The scheme mainly uses attribute-based encryption and dynamic proactive secret sharing to control redacting rights, which is improved in public accountability compared with the previous scheme. In the same year, Panwar et al. \cite{panwar2021retrace} proposed a framework for effective rewriting of blockchain content, which uses the method based on chameleon hash and dynamic group signature. However, these two schemes use complex cryptographic tools, which will bring high overhead.
In addition, Xu et al.\cite{xu2021k} proposed a redactable blockchain scheme with money punishment mechanism. In this scheme, the authority CA stipulates that redactor can only modify \emph{k} times at most, and each redactor needs to place a time lock deposit on the chain. If the redactor acts maliciously within the specified time, then CA can take the deposit as punishment. However, if the redactors who meet the policy make malicious modifications to the transaction content within the legal \emph{k} times, other users cannot intervene in these redacting operations, which will be detrimental to data supervision.

\subsection{Motivation}
In the above two redacting consensus, the redacting consensus based on the chameleon hash function is more applicable to the real world than the other. This is because the use of the chameleon hash function will not change the hash value, which can naturally ensure the integrity of the hash chain, and the redacting efficiency is superior. It can achieve timely redacting of data without waiting as long as the voting based method. However, there are still many problems in the current redactable blockchain scheme based on chameleon hash function.
First, some schemes use complex cryptographic tools such as secure multi-party computation and secret sharing, which will bring high communication overhead and reduce redacting efficiency. Secondly, when trapdoors are given to users, there is a lack of corresponding trapdoor revocation mechanism, which will make the transaction owner unable to revoke the redacting rights of some users, and malicious redactors can modify the transaction indefinitely, or conspire with users who do not meet the redacting policy to tamper with the transaction.
Thirdly, many schemes are coarse-grained redacting, which means that they can only be modified in blocks, lacking transaction level fine-grained redaction. Finally, the transaction owner may set a policy that only allows himself to redact the transaction, so that others cannot prevent the malicious owner from modifying the transaction arbitrarily, lacking a regulatory mechanism.
Motivated by the above problems, we try to design appropriate solutions to achieve the following goals: (1) Efficient redacting consensus: it only takes a short time to complete a redacting operation to achieve the function of timely redacting. (2) Updateable policy: after setting a policy, the transaction owner can revoke or add rules in the policy in time to control the range of legal redactors who meet the rules. (3) Transaction level redacting: changing a transaction does not require the participation of other transactions to reduce the range of redacting operations.
(4) Public accountability and supervision mechanism: it can prevent redactors from abusing redacting rights and maliciously tampering with transactions to ensure the security of the blockchain.

\subsection{Our Contributions}
The main aim of our scheme are the above four. Specifically, the contribution of this paper is mainly divided into the following three aspects:

\begin{enumerate}[(1)]
  \item Firstly, we introduced the concept of a witness group as an important part of the accountability and supervision mechanism. The witness group can prevent malicious redactors from redacting transactions arbitrarily.
  \item Secondly, we propose an efficient and fine-grained redactable blockchain scheme with accountability and updatable policies, called \emph{EFRB}. In our scheme, the transaction owner can set a set of updatable policies associated with the transaction. Only users who meet the policies can become the legal redactor of the transaction.
  Our scheme mainly uses chameleon hash function and digital signature, and makes the trapdoor of chameleon hash function public, so it can avoid many security problems caused by trapdoor management.
  The key to a successful redacting operation is whether the transaction has been signed by both the legal redactor and the witness group. The use of digital signature can not only achieve efficient redacting consensus, but also identify the redactor and the witness group responsible for the transaction through signature, so as to achieve accountability and punishment mechanisms.
  \item Finally, we implemented \emph{EFRB} scheme, and carried out a series of experiments and analysis. The results show that our scheme can achieve efficient and fine-grained redacting functions with accountability and updatable policies.
\end{enumerate}

The rest of the article is organized as follows: In Section \ref{s2}, we introduce some preliminary knowledge involved in \emph{EFRB} scheme, including digital signature, chameleon hash function and policy. In Section \ref{s3}, we introduce the concept of witness group and give the method of selecting witness group.
In Section \ref{s4}, we propose the general framework of \emph{EFRB} and give the details of the scheme, including security model and threat model. In Section \ref{s5}, we implemented the \emph{EFRB} scheme, carried out a series of experiments, and analyzed the security and performance of the scheme. In Section \ref{s6}, we summarize this article.

\section{Preliminaries}
\label{s2}
In this section, we mainly introduce some necessary preliminary knowledge involved in \emph{EFRB} scheme, including chameleon hash function, digital signature and policy.

\subsection{Blockchain}
In this subsection, we refer to the abstract way of describing blockchain in \cite{garay2015bitcoin,david2018ouroboros}, and review the basic definition of blockchain.
A block $\mathfrak{B}_{i}:=(sl_{i},ph_{i},mt_{i},ne_{i},\mathcal{T}_{i})$ is a quintet form. The first four elements are the head of the block and the last one is the body of the block. Here, $sl_{i}$ is the discrete time unit slots executed by the blockchain system, $ph_{i}$ is the hash value of the previous block header,
$mt_{i}$ is the root of the Merkle Tree composed of transactions $\mathcal{T}_{i}$ in the block, $ne_{i}$ is the nonce of PoW consensus algorithm. The transaction set contains many transactions, that is, $\mathcal{T}_{i}=(tx_{1},tx_{2},\cdots,tx_{m})$, where \emph{m} is the maximum number of transactions.

Blockchain $\mathfrak{C}$ is composed of a string of the above blocks, that is, $\mathfrak{C}:=(\mathfrak{B}_{1},\mathfrak{B}_{2},\cdots,\mathfrak{B}_{l})$, where \emph{l} is the length of a chain $\mathfrak{C}$. The leftmost block $\mathfrak{B}_{1}$ in the chain is the genesis block,
and the rightmost block $\mathfrak{B}_{l}$ is the head of the blockchain at time \emph{sl}, which is denoted by $Header_{sl}(\mathfrak{C}):=\mathfrak{B}_{l}$. Any chain can be expanded into a new chain $\mathfrak{C}':=\mathfrak{C}||\mathfrak{B}_{l+1}$ through some consensus mechanisms, where $\mathfrak{B}_{l+1}:=(sl_{l+1},ph_{l+1},mt_{l+1},ne_{l+1},\mathcal{T}_{l+1})$.

In terms of consensus algorithm, traditional blockchains such as Bitcoin mainly use PoW algorithm to run for block producers. In short, the algorithm is to find a random value \emph{ne} so that the hash value of the data in the new block header is less than a given value \emph{T},
i.e. $\mathcal{H}(sl,ph,mt,ne)<T$, where $\mathcal{H}$ is the hash functions in cryptography, which satisfies the collision-resistant property. In short, it is almost impossible to find a pair of different messages \emph{m} and \emph{m'}, making $\mathcal{H}(m)=\mathcal{H}(m')$.

\subsection{Chameleon Hash Function}
\newtheorem{Def}{Definition}
\begin{Def}
  (Chameleon Hash Function) Generally speaking, a chameleon hash function $\mathcal{C} \mathcal{H}$ is mainly composed of the following five effective algorithms:\\
  \indent $PGen_{\mathcal{C} \mathcal{H}}(1^{\kappa})$. The algorithm takes a security parameter $\kappa$ as the input, where $\kappa$ belongs to a set $\mathbb{N}$ composed of all natural numbers, outputs a common parameter pp, and takes it as an implicit input of other algorithms.\\
  \indent $KGen_{\mathcal{C} \mathcal{H}}(pp)$. The algorithm takes the public parameter pp as the input and outputs a pair of public and secret keys (pk,sk).\\
  \indent $Hash_{\mathcal{C} \mathcal{H}}(pk,m)$. The algorithm takes a public key pk and a message m as inputs, where message m belongs to a message space $\mathcal{M}$, and outputs a random number r and a hash value h.\\
  \indent $Ver_{\mathcal{C} \mathcal{H}}(pk,h,r,m)$. The algorithm takes a public key pk, a message m, a hash value h and a random number r as inputs, and outputs 0 or 1, where 1 and 0 represent success and failure of verification respectively.\\
  \indent $Apt_{\mathcal{C} \mathcal{H}}(sk,h,r,m,m')$. The algorithm takes a secret key sk, a hash value h, a random number r, an old message m and a new message $m'$ as inputs, and outputs a new random number $r'$.
\end{Def}

\begin{Def}
  (Correctness) If for all security parameter $\kappa \in \mathbb{N}$, for all public parameter $pp\leftarrow PGen_{\mathcal{C} \mathcal{H}}(1^{\kappa})$, for all key pairs $(pk,sk)\leftarrow KGen_{\mathcal{C} \mathcal{H}}(pp)$, for all message $m,m'\in \mathcal{M}$,
  for all $(r,h)\leftarrow Hash_{\mathcal{C} \mathcal{H}}(pk,m)$, we have for all $r'\leftarrow Apt_{\mathcal{C} \mathcal{H}}(sk,h,r,m,m')$, that $Ver_{\mathcal{C} \mathcal{H}}(pk,h,r,m)=Ver_{\mathcal{C} \mathcal{H}}(pk,h,r',m')=1$ is always true, then the chameleon hash function $\mathcal{C} \mathcal{H}$ is called correct.
\end{Def}

Note that in our scheme, the chameleon hash function is used to provide the function of modifying data and maintaining the integrity of the blockchain. Because its trapdoor is public, everyone can use the trapdoor to redact the transaction content,
but only if the witness group and the legitimate redactor who meets the transaction policy sign the redacted transaction together can the redacting be considered successful. Therefore, the security of the scheme is not based on the chameleon hash function, but is guaranteed by the security of the digital signature.

\subsection{Policy}
A policy is a set of rules. In the redactable blockchain, it is used to restrict redacting rights, that is, the transaction owner can specify who can redact the transaction. For subsequent explanation, we set $\mathcal{M}(\mathcal{P},\mathcal{S})$ to represent the matching function of attribute set $\mathcal{S}$ and policy $\mathcal{P}$. If the result is \emph{1}, $\mathcal{S}$ can match $\mathcal{P}$, and if it is \emph{0}, it means that the matching fails.
The current redactable blockchain scheme mainly uses attribute-based encryption (ABE) \cite{goyal2006attribute}, which is divided into ciphertext-policy attribute-based encryption (CP-ABE) \cite{bethencourt2007ciphertext} and key-policy attribute-based encryption (KP-ABE) \cite{han2012privacy}.
To be specific, CP-ABE means that users can encrypt some information and embed the policies they want to set into the ciphertext, and other users' attributes are embedded into their own keys. When the attributes of other users can match the policy, they can decrypt the ciphertext with their own keys.
For example, user $u_{1}$ encrypts a message to get ciphertext \emph{c}, and sets the policy $\mathcal{P}=$("Teacher" OR "Student"), user $u_{2}$ has the attribute $S_{2}=$("Student"), and user $u_{3}$ has the attribute $S_{3}=$("Salesman"). Since $\mathcal{M}(\mathcal{P},\mathcal{S}_{2})=1$, $u_{2}$ can decrypt \emph{c} with his secret key,
while $u_{3}$ cannot decrypt \emph{c} with his secret key because $\mathcal{M}(\mathcal{P},\mathcal{S}_{3})=0$. KP-ABE and CP-ABE set the policy in the opposite way. In KP-ABE, the policy is embedded in the key, while the attribute is embedded in the ciphertext. Inspired by the idea of CP-ABE, we do not use specific ABE scheme, but let a central authority (CA) authenticate the attributes of users and sign the set consisting of attributes and user's public key after authentication.

\subsection{Digital Signature}
\begin{Def}
  (Digital signature) A digital signature $\mathcal{D} \mathcal{S}$ is mainly composed of the following four effective algorithms:\\
  \indent $PGen_{\mathcal{D} \mathcal{S}}(1^{\kappa})$. The algorithm takes a security parameter $\kappa$ as the input, where $\kappa\in \mathbb{N}$, outputs a common parameter pp, and takes it as an implicit input of other algorithms.\\
  \indent $KGen_{\mathcal{D} \mathcal{S}}(pp)$. The algorithm takes the public parameter pp as the input and outputs a pair of public and secret keys (pk,sk).\\
  \indent $Sig_{\mathcal{D} \mathcal{S}}(m,sk)$. The algorithm takes a message $m\in \mathcal{M}$ and a secret key sk as inputs, and outputs a signature $\sigma$.\\
  \indent $Ver_{\mathcal{D} \mathcal{S}}(\sigma,m,pk)$. The algorithm takes a signature $\sigma$, a message m and a public key pk as inputs, and outputs 0 or 1, where 1 and 0 represent success and failure of verification respectively.
\end{Def}

\begin{Def}
  (Correctness) If for all security parameter $\kappa \in \mathbb{N}$, for all public parameter $pp\leftarrow PGen_{\mathcal{D} \mathcal{S}}(1^{\kappa})$, for all key pairs $(pk,sk)\leftarrow KGen_{\mathcal{D} \mathcal{S}}(pp)$, for all message $m\in \mathcal{M}$,
  $Ver_{\mathcal{D} \mathcal{S}}(Sig_{\mathcal{D} \mathcal{S}}(m,sk),m,pk)=1$ is always true, then the digital signature $\mathcal{D} \mathcal{S}$ is called correct.
\end{Def}

In addition to being correct, a secure digital signature should also be unforgeable. That is, except for the negligible probability, no adversary can generate a signature for the message $m^{*}$ that has never been seen.
A function $\varepsilon (\kappa)$ is said to be negligible if for all polynomials $p(\kappa)$ and all sufficiently large $\kappa\in \mathbb{N}$, it has $\varepsilon (\kappa)<1/p(\kappa)$. In our \emph{EFRB} scheme, the security of digital signature $\mathcal{D} \mathcal{S}$ needs to satisfy the existential unforgeability under adaptively chosen-message attack (EUF-CMA).

\begin{Def}
  (EUF-CMA) If for all probabilistic polynomial time (PPT) adversary $\mathcal{A}$ having access to a signing oracle $\mathcal{O}_{Sig}(sk,\cdot)$, the following advantage of $\mathcal{A}$ are negligible, then the digital signature $\mathcal{D} \mathcal{S}$ has EUF-CMA security.\\
  $Pr[pp\leftarrow PGen_{\mathcal{D} \mathcal{S}}(1^{\kappa});(pk,sk)\leftarrow KGen_{\mathcal{D} \mathcal{S}}(pp);(m^{*},\sigma ^{*})\leftarrow \mathcal{A}^{\mathcal{O}_{Sig}(sk,\cdot)}(pk):m^{*}\neq m \forall m\in \mathcal{Q} \wedge Ver_{\mathcal{D} \mathcal{S}}(\sigma ^{*},m^{*},pk)=1]=\varepsilon (\kappa)$, where $\mathcal{Q}$ is the set of queries which $\mathcal{A}$ made to $\mathcal{O}_{Sig}(sk,\cdot)$.
\end{Def}

\section{Witness Group}
\label{s3}
In this section, we will introduce the concept of witness group, and how to select and use witness group.

\subsection{Concept of Witness Group}
Because in our scheme, the trapdoor of the chameleon hash function is open, which means that everyone can use the trapdoor to redact transactions, but only those users whose attribute sets meet the transaction policy are legal redactors, and a redacted transaction requires the digital signature of the legal redactor to be an effective redacting operation.
However, it is not enough to only use the redactor's signature, because the transaction owner may have set a transaction policy that can only match his own attribute set, so that only he can redact the transaction, which may lead to the redactor abusing his redacting rights.
To solve this problem, we introduce a concept called "witness group", in which users will act as supervisors of redacting operations. That is, when and only when a redacted transaction has the digital signatures of both the legal redactor and the witness group, the redacted transaction is valid and will be recognized by other users.

\subsection{Selection of Witness Group}
How to select the witness group is a key issue, because it is related to the security of the scheme and whether the witness group is useful for redacting consensus. A good witness group should be trusted by most users of the blockchain system, so the witnesses selected in some way should be the honest majority.
Inspired by \cite{gilad2017algorand} and \cite{li2021escaping}, because the underlying consensus of the blockchain system uses the amount of wealth held by users, such as computing power in PoW and stake in PoS, users with more wealth are more likely to maintain the blockchain system. Therefore, we consider selecting witnesses according to the wealth value held by users, and the algorithm for selecting witness group denoted by \emph{Sel}.
In addition, because different systems have different wealth units, the selection methods will be different. For example, in PoS, wealth is a stake. We can use verifiable random function (VRF) \cite{micali1999verifiable} to select witnesses, and use the return value of the function to obtain the weight of witnesses.
In PoW, wealth is computing power. We can reduce the difficulty of underlying consensus to select witnesses, and take the number of puzzles solved as the voting weight. Next, we will take the PoW system as an example to give the specific algorithm of \emph{Sel}. It should be noted that the specific cycle and duration of witness selection are independent of the underlying block generation time, and are mainly determined by the network environment and system parameters.

\begin{algorithm}
  \footnotesize
  \caption{Witness Selection Algorithm \emph{Sel}}
  \label{Sel}
  \begin{algorithmic}[1]
    \REQUIRE 
      block header $bh:=(sl,ph,mt,ne)$, moderate target value \emph{tv}, select period \emph{sp}, user public key \emph{pk}
    \ENSURE
      weight \emph{w} and proof \emph{info}
    \STATE $w:=0$, $info:=\varnothing$, \emph{tm} is the current time;
    \WHILE {$tm\in sp$}
    \IF {find \emph{ne} to make $\mathcal{H}(tm,ph,mt,ne,pk)<tv$}
    \STATE $w=w+1$, $info=info\cup (tm,ph,mt,ne,pk)$;
    \ENDIF
    \ENDWHILE
    \RETURN $(w,info)$;
  \end{algorithmic}
\end{algorithm}

\indent Algorithm \emph{VSel} is the corresponding algorithm to verify whether the witness is legal.

\begin{algorithm}
  \footnotesize
  \caption{Verifying Sel Algorithm \emph{VSel}}
  \label{VSel}
  \begin{algorithmic}[1]
    \REQUIRE 
      weight \emph{w}, proof \emph{info}, select period \emph{sp}
    \ENSURE
      0 or 1
    \STATE parse $info:=(tm,ph,mt,ne,pk)$, $c:=0$;
    \FOR {each \emph{info}}
    \IF {$tm\in sp\wedge \mathcal{H}(m,ph,mt,ne,pk)<tv\wedge ph=\mathcal{H}(Header_{tm}(\mathfrak{C}))$}
    \STATE $c=c+1$;
    \ENDIF
    \ENDFOR
    \IF {$c=w$}
    \RETURN 1;
    \ENDIF
    \RETURN 0;
  \end{algorithmic}
\end{algorithm}  

Because selecting witness group requires a certain amount of time and overhead, it is unnecessary to select witnesses frequently. The system only needs to dynamically set a period of time before reselecting witness group, such as once a day.

\subsection{Use of Witness Group}
After the witness group is selected, the user with the largest weight will become a voting collector. If the collector has bad behavior, such as no response, lost votes, etc., a new collector will be selected from the witness group according to the weight ranking or random selection.
When the witness group members receive a redacting request, they vote on the redacting request and send the signed vote to the collector. Only when the cumulative voting weight exceeds a certain threshold can it be considered that the witness group has agreed to the redacting request.
After that, the collector aggregates these signatures and sends it to other nodes. When other users receive a redacted transaction with the signature of the redactor and witness group, verify its legitimacy. If the verification is passed, the user will replace the original transaction with a new transaction to update their local blockchain.

Although it is ensured that the selected witnesses are the honest majority, the selected witnesses may also include malicious users, so users need to supervise each other. That is to say, the witnesses need to pay a certain amount of security deposit,
and the deposit can be withdrawn only after leaving the witness group for a period of time. When a transaction is maliciously redacted and signed by the redactor and witness group, the user who finds the behavior can report it to the CA.
Because there are signatures of witness groups in redacting transactions, they can be identified publicly and punished to achieve accountability. As a punishment, the CA will deduct the deposits of all witnesses responsible for the transaction and select a new witness group.
As a reward, the discoverer can be rewarded from the witness's deposit. In addition, in order to encourage honest users to actively participate in witness selection, we require an additional processing fee for each redacting request.
After witnesses leave the witness group and get back the deposit, they will be rewarded with additional processing fees. If the redacting request fails or the witness group is reported, the processing fee will be returned to the redactor.

In addition, it is noted that the \emph{Sel} algorithm and the underlying consensus algorithm are parallel, which will not affect the underlying consensus mechanism. That is, users may meet the conditions for becoming witnesses when running the underlying consensus.
For example, in PoW, the essence of the \emph{Sel} algorithm is to reduce the difficulty of puzzles in the underlying consensus. Therefore, users may get the nonce of puzzles in the \emph{Sel} algorithm during the process of participating in the election for the block producer, but it depends on whether users want to be witnesses.

In general, we divide the participation of the witness group in redacting consensus into four parts: (1) The witness group is randomly selected through the algorithm \emph{Sel} and the wealth value held by the user.
(2) The members of the witness group jointly decide whether to sign the redacted transaction. (3) The system periodically selects witness group. (4) The punishment and incentive mechanism of joint and several liability.

\section{Efficient and Fine-grained Redactable Blockchain Supporting Accountability and Updating Policies}
\label{s4}
In this section, we will show the details of the proposed \emph{EFRB} scheme, including \emph{EFRB}'s general framework, threat model and security model.

The roles in our scheme are composed of users and central authority (CA) of the blockchain system, in which users are divided into ordinary users, transaction owners, redactors, witness group and block producer. Ordinary users are users without the latter three identities.
It should be noted that these user classifications are not independent, that is, a user may have multiple identities, for example, he may be both a transaction owner and a redactor.

\begin{itemize}
 \item As an authority, CA is trusted by users and acts as a blockchain administrator. CA is mainly responsible for authenticating user attributes and issuing attribute certificates for users. In addition, it is also responsible for handling the information reported, that is, when users find that witness groups abuse their signature rights, they report this behavior to the CA,
 and the CA will deduct the witness's deposit after verification, select a new witness group, and reward the reporter with the deposit.
 \item The transaction owner is the sender of a transaction, who can choose whether to make the transaction redactable. If yes, he needs to use $\mathcal{C} \mathcal{H}$ to calculate the hash of the transaction and set the redacting policy of the transaction, that is, specify which attributes the legal redactor of the transaction needs to have,
 and then attach the trapdoor, policy and other elements of $\mathcal{C} \mathcal{H}$ to the transaction.
 \item Redactors are users whose own attribute sets conform to the transaction policy, which means they have the right to redact transactions. Therefore, if a transaction needs to be redacted, they can redact and sign the transaction and make a redacting request to the witness group.
 \item The witness group is the key to the scheme. It is formed by selecting witnesses from users according to the algorithm \emph{Sel}. They are responsible for reviewing redacting requests and voting to decide whether to approve the redacting operation. Those approved redacting transactions will be signed by the witness group.
 \item Block producers are miners who generate blocks. In our scheme, his main task is to package the information of transactions and witness groups together into new blocks, update the blockchain, and make the whole network know which users comprise the current witness group.
\end{itemize}

\subsection{Overview of EFRB}
\begin{figure}\centering \includegraphics[width=0.4\textwidth]{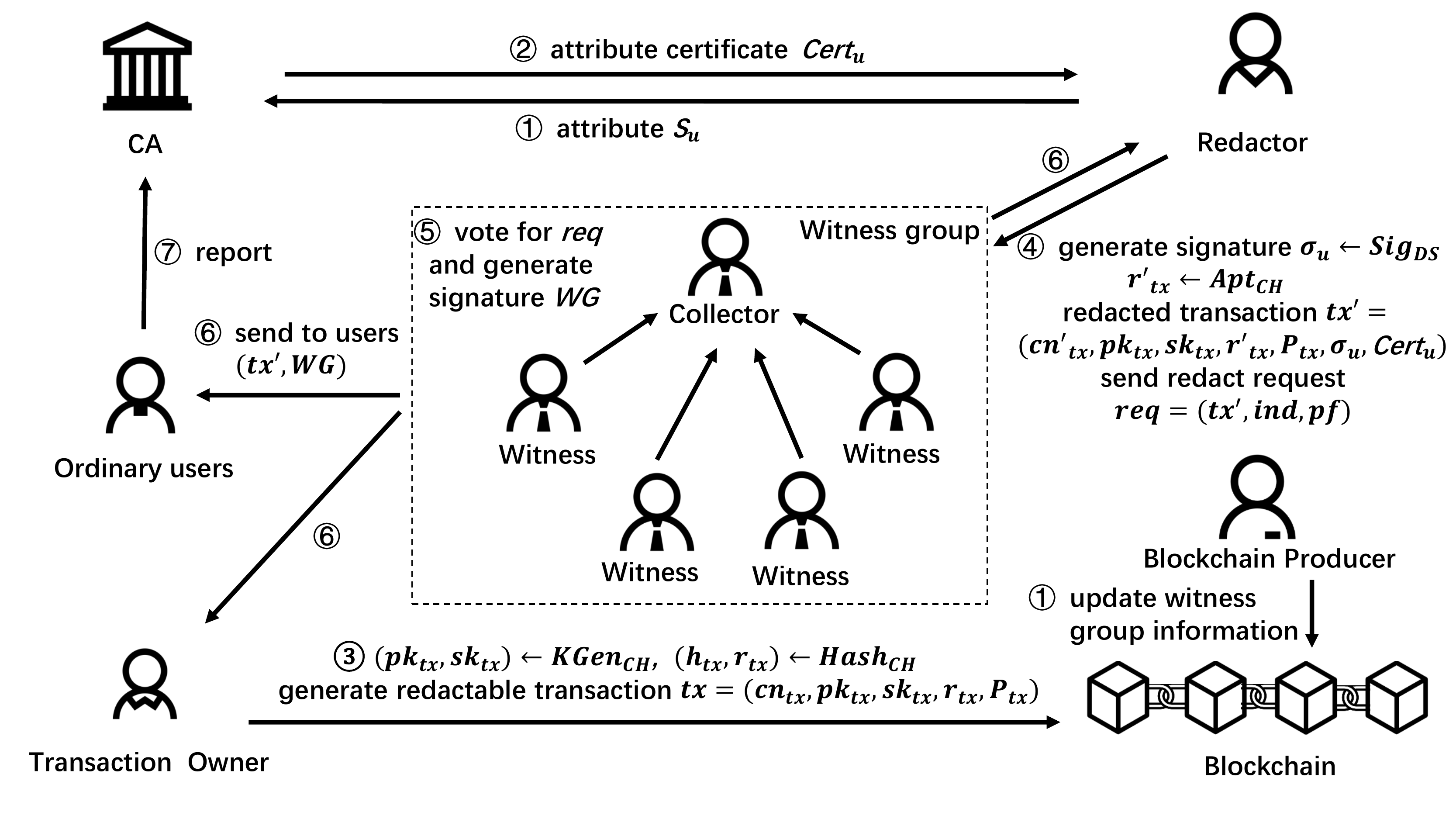}  \caption{EFRB scheme diagram} \vspace{-0.2in} \label{fig1} \end{figure}

Our \emph{EFRB} scheme is shown in Figure \ref{fig1}. The main idea of the scheme is: First, the user can send a set of his own attributes $\mathcal{S}_{u}$ to the CA. After CA authentication, the user will obtain an attribute certificate $Cert_{u}$, including the user's attribute $\mathcal{S}_{u}$, the user's public key $pk_{u}$, and the CA's signature $\sigma_{ca}$ for these data.
Then, the transaction owner can choose whether the transaction \emph{tx} is redactable or immutable. If it is immutable, it will process \emph{tx} in the same way as the traditional blockchain in the future. If it is redactable, it will use $\mathcal{C} \mathcal{H}$ to generate a hash value of \emph{tx}, and put the public key $pk_{tx}$, trapdoor $sk_{tx}$, random number $r_{tx}$ of $\mathcal{C} \mathcal{H}$, and the transaction redacting policy $\mathcal{P}_{tx}$ into \emph{tx}.

When \emph{tx} needs to be redacted, the user who meets the attributes in $\mathcal{P}$ will become the redactor, and he will perform redacting operations to generate the redacted new transaction \emph{tx'}. These operations include updating the transaction content \emph{tx}, using the $sk_{tx}$ in the transaction to update the random number $r_{tx}$ to $r_{tx}'$,
attaching his own attribute certificate $Cert_{u}$ to the \emph{tx'}, and signing the redacted transaction. Note that random numbers, attribute certificates, and signatures themselves are not included in the hash function. Later, he will send an redacting request \emph{req} to the current witness group of the system, which consists of the redacted transaction \emph{tx'}, the index \emph{ind} where the transaction is located, and the processing fee \emph{pf}.

When the witness group receives the redacting request \emph{req}, they will reviews its content $cn_{tx}$ and verify whether the identity of the redactor is legal, that is, whether $\mathcal{M}(\mathcal{P},\mathcal{S}_{u})=1$. Then, they vote in the group. If they agree to the redacting request, they sign the hash value of \emph{tx'} and send it to the collector. When the cumulative voting weight reaches a certain threshold \emph{ts},
the collector aggregates these signatures into \emph{WG}, appends it to \emph{tx'}, and then sends the signed transaction request to other users.

Finally, when other users receive the transaction \emph{tx'} with the signature of the redactor and witness group, verify its legitimacy. If the verification is successful, users will replace the original transaction \emph{tx} of the blockchain with the new transaction \emph{tx'} to update their local blockchain.
If the transaction owner wants to modify policy $\mathcal{P}$ in the future, he can send a redacting request, and other nodes will process it in the same way as described above.

\subsection{Formal Description of EFRB} %system model part
In this subsection, we will describe the \emph{EFRB} scheme in detail, which is mainly composed of four parts.

\paragraph{\textbf{Initialization}} All users and CA need to participate in initialization, mainly including the following three parts:
\begin{enumerate}[1)]
  \item \emph{System parameter initialization}. CA generates public parameters $pp_{\mathcal{D} \mathcal{S}}$ and $pp_{\mathcal{C} \mathcal{H}}$ of digital signature $\mathcal{D} \mathcal{S}$ and chameleon hash function $\mathcal{C} \mathcal{H}$ through $PGen_{\mathcal{D} \mathcal{S}}(1^{\kappa})$ and $PGen_{\mathcal{C} \mathcal{H}}(1^{\kappa})$ algorithms respectively, and broadcasts them to other users.
  \item \emph{User key pair initialization}. After obtaining $pp_{\mathcal{D} \mathcal{S}}$ and $pp_{\mathcal{C} \mathcal{H}}$, each user, including CA, uses the $KGen_{\mathcal{D} \mathcal{S}}(pp_{\mathcal{D} \mathcal{S}})$ algorithm to generate their own public and secret key pair $(pk_{u},sk_{u})$, and then makes the $pk_{u}$ public, where the key pair of CA is $(pk_{ca},sk_{ca})$.
  \item \emph{Attribute set authentication for the user}. When a user wants to authenticate his attribute set $\mathcal{S}_{u}$, he can send $\mathcal{S}_{u}$ and his public key $pk_{u}$ to the CA. After receiving the authentication request, the CA reviews the user's identity and attribute set. After the authentication is successful,
  the CA will use his own secret key $sk_{ca}$ and the $Sig_{\mathcal{D} \mathcal{S}}$ algorithm to sign the user's public key $pk_{u}$ and attribute set $\mathcal{S}_{u}$, generate an attribute certificate $Cert_{u}=(\sigma _{ca},(pk_{u},\mathcal{S}_{u}))$, and send it to the user.
\end{enumerate}

\paragraph{\textbf{Generate redactable transactions}} When generating a transaction, the transaction owner can set the transaction type as immutable or redactable. When the transaction is immutable, it will be handled in the same way as the traditional blockchain in the future. When the transaction is redactable, the transaction owner will perform the following operations to generate a redactable transaction:
\begin{enumerate}[1)]
  \item \emph{Generate $\mathcal{C} \mathcal{H}$ keys}. The transaction owner first needs to use the public parameter $pp_{\mathcal{C} \mathcal{H}}$ and the algorithm $KGen_{\mathcal{C} \mathcal{H}}(pp_{\mathcal{C} \mathcal{H}})$ to generate a pair of public and secret keys $(pk_{tx},sk_{tx})$ of $\mathcal{C} \mathcal{H}$.
  \item \emph{Set Policy}. The transaction owner needs to set a transaction policy $\mathcal{P}_{tx}$, in which the user with the specified attribute can have the right to redact the transaction.
  \item \emph{Attach the necessary elements to the transaction}. The algorithm $Hash_{\mathcal{C} \mathcal{H}}$ is used to obtain transaction hash value $h_{tx}$ and a random number $r_{tx}$. The transaction owner needs to attach the public and secret keys pair $(pk_{tx},sk_{tx})$ of $\mathcal{C} \mathcal{H}$, random number $r_{tx}$, and policy $\mathcal{P}_{tx}$ to the transaction content $cn_{tx}$ to generate a redactable transaction $tx=(cn_{tx},h_{tx},pk_{tx},sk_{tx},r_{tx},\mathcal{P}_{tx})$ and send it to other users. After that, the process of adding the transaction to the blockchain is the same as that of traditional blockchains.
\end{enumerate}

\paragraph{\textbf{Select witness group}} This part is the key of \emph{EFRB} scheme, which is mainly divided into the following modules. We have discussed the details in the section \ref{s3} of this article.
\begin{enumerate}[1)]
  \item \emph{Periodically select witness group}. If users want to be witnesses, they can use the algorithm \emph{Sel} to campaign within the specified period to obtain the voting weight \emph{w} and the corresponding proof information \emph{info}, and send them to other users. Assuming that the system selects a witness group every \emph{t} slots, the duration of each election is \emph{sp}, and the difficulty of the puzzles in the current \emph{Sel} algorithm is \emph{tv}, then for the current time \emph{sl}, the current blockchain head $Header_{sl}(\mathfrak{C})=(sl,ph,mt,ne,\mathcal{T})$, and the system has elected \emph{m} witness groups so far.
  If $sl\in [t*m,t*m+sp]$, the system starts the election of the \emph{m+1} witness group, and each user \emph{u} executes $(w,info)\leftarrow Sel(sl,ph,mt,ne,tv,sp,pk_{u})$. Other users verify the validity of the results through the algorithm $VSel(w,info)$. Then the block producer will take the first few of the received legal results with greater weight as the witness group, package the results and ordinary transactions together into a new block, and broadcast the new block to other users.
  \item \emph{Selection of collectors}. The user with the largest weight will become the voting collector in the witness group, and he will locally record the corresponding relationship $\mathcal{F}(pk)=w$ between the public key and weight of legal witnesses. If the collector has bad behavior, it will be sorted in descending order according to the weight of the witness and select other witnesses from the sorting results to replace him.
  \item \emph{Rewards and punishments}. When users become witnesses, they need to pay a deposit to CA to prevent them from abusing voting right. When a witness group $G_{1}$ is normally updated to $G_{2}$, if $G_{1}$ is not reported by other users during the period when $G_{1}$ is a witness group and within a period of time after the witness group is updated to $G_{2}$, the witnesses in the $G_{1}$ can get back the deposit and receive additional rewards. This reward comes from the redacting fees in those redacting requests that they processed as witnesses. However, if $G_{1}$ is reported by users and confirmed by CA, they will be deducted all the deposits and rewarded to the reporter.
\end{enumerate}

\paragraph{\textbf{Redact transaction}} When a transaction needs to be redacted, it can be divided into the following steps:
\begin{enumerate}[1)]
  \item \emph{Make a redacting request}. If a user wants to redact a transaction $tx=(cn_{tx},h_{tx},pk_{tx},sk_{tx},r_{tx},\mathcal{P}_{tx})$, he first needs to use $\mathcal{M}(\mathcal{P}_{tx},\mathcal{S}_{r})$ to check whether his attributes $\mathcal{S}_{r}$ match the policy $\mathcal{P}_{tx}$ of the transaction. If so, the user is the redactor of the transaction and has the right to redact the transaction. Secondly, the redactor changes the old transaction content $cn_{tx}$ to the new content $cn_{tx}'$, updates the random number $r_{tx}$ to $r_{tx}'$ using the trapdoor $sk_{tx}$ and algorithm $Apt_{\mathcal{C} \mathcal{H}}$,
  and adds his attribute certificate $Cert_{r}=(\sigma _{ca},(pk_{r},\mathcal{S}_{r}))$ to the \emph{tx}. Thirdly, the redactor uses the algorithm $Sig_{\mathcal{D} \mathcal{S}}(tx',sk_{r})$ to sign the redacted transaction and append the signature $\sigma _{r}$ to the transaction \emph{tx'}. Finally, the redactor puts forward the redacting request \emph{req} and sends it to the witness group. The \emph{req} includes the redacted transaction $tx'$, the index of the transaction in the blockchain \emph{ind}, and the processing fee \emph{pf} for redacting the transaction.
  \item \emph{Consensus among witnesses}. When the witness group receives the redacting request, each witness reviews the request, including using $Ver_{\mathcal{D} \mathcal{S}}$ algorithm to verify the identity of the redactor and using $Ver_{\mathcal{C} \mathcal{H}}$ algorithm to verify the legitimacy of the redacting content. If the verification is passed and the witness agrees to the redacting request, he can send a vote \emph{v} to the collector in the group, including his signature of the redacted transaction using $Sig_{\mathcal{D} \mathcal{S}}(tx',sk_{w})$ and his public key $pk_{w}$.
  After the collector receives the vote, the $Ver_{\mathcal{D} \mathcal{S}}$ is used to verify the legitimacy of the vote, and the weight of the voter is obtained according to the public key. When the cumulative weight exceeds a certain threshold \emph{ts}, it means that the whole witness group reaches a consensus on the \emph{req}. The collector will aggregate these signatures, append them to the redacted transaction, and then broadcast the signed redacting request to other users.
  \item \emph{Update transaction}. Other users verify the validity of redacting requests signed by redactors and witnesses through $Ver_{\mathcal{D} \mathcal{S}}$, $Ver_{\mathcal{C} \mathcal{H}}$ and $VSel$ algorithms. If the verification is passed, users will update the redacted transaction to the original transaction location.
\end{enumerate}

If the transaction owner wants to modify the transaction policy, he can also redact the policy through the above methods.

\subsection{Threat Model}
In the \emph{EFRB} scheme, we assume that CA is a third party that is fully trusted by all users of the system. Since the witness group is selected by a random algorithm \emph{Sel}, which can ensure that the selected people are the honest majority, but there may still be malicious users, the witness group is semi-trusted. As the maintainer of the blockchain, the block producer is also trusted by most users. Ordinary users, transaction owners and transaction redactors are untrusted.
They may jointly launch collusive attacks with malicious users in the witness group. The following threats may exist in our scheme: (1) Users who do not meet the transaction policy try to modify the transaction. (2) The policy set by the transaction owner only allows him to redact the transaction. (3) Malicious witnesses abused their voting rights, or voted for transactions that were maliciously redacted, and tried to escape punishment.

Since we do not limit the number of times the redactor can redact a transaction, that is, the redactor is allowed to redact the transaction indefinitely, unless the transaction owner has updated the transaction policy. However, \cite{xu2021k} only considers the threat model in the \emph{K} times case, and does not consider the second threat scenario above. In our scheme, we mainly use the unforgeability of digital signature and the collision-resistant property of hash function to ensure the security of the system.
Therefore, even if the trapdoor of the chameleon hash function is disclosed, if the signature of any party of the redactor or witness group is missing from the redacted transaction, the redacted transaction is invalid. Because the attribute certificate of the redactor is granted by the CA, the witness group is selected through the \emph{Sel} algorithm, which refers to the thought of safely selecting committee members in \cite{li2021escaping}.
Like the underlying consensus mechanism of the blockchain, it can ensure that the selected witnesses are the honest majority. That is, if the secret key of the CA is confidential and the proportion of adversaries among all users of the system does not exceed $1/2$, then our scheme is secure.

\subsection{Security Model}
According to the above threat model, we give the following security model.
\begin{Def}
  (EUF-CMA Security) Assume that there are several oracles as follows: a hash oracle $\mathcal{O}_{1}$, a redact oracle $\mathcal{O}_{2}$, a redactor initialization oracle $\mathcal{O}_{3}$, a redactor corrupt oracle $\mathcal{O}_{4}$, a user initialization oracle $\mathcal{O}_{5}$, a user corrupt oracle $\mathcal{O}_{6}$, a attribute certificate generation oracle $\mathcal{O}_{7}$, a witness initialization oracle $\mathcal{O}_{8}$, and a witness corrupt oracle $\mathcal{O}_{9}$.
  Then the security of \emph{EFRB} is based on the following experiment, that is, if for any PPT adversary $\mathcal{A}$, $Pr[Exp^{EUF-CMA}_{EFRB,\mathcal{A}}(1^{\kappa})]=1$ is negligible, then we call \emph{EFRB} is EUF-CMA secure.
\end{Def}

\begin{table*}\normalsize
  \renewcommand\arraystretch{1.2}
  \centering
  \begin{tabular}{|p{\textwidth}|}
    \hline
    $Exp^{EUF-CMA}_{EFRB,\mathcal{A}}(1^{\kappa})$\\
    $pp_{\mathcal{D} \mathcal{S}}\leftarrow PGen_{\mathcal{D} \mathcal{S}}(1^{\kappa})$, $pp_{\mathcal{C} \mathcal{H}}\leftarrow PGen_{\mathcal{C} \mathcal{H}}(1^{\kappa})$, $((pk_{ca},pk_{r},pk_{u},pk_{w},pk_{tx}),(sk_{ca},sk_{r},sk_{u},sk_{w},sk_{tx}))\leftarrow KGen_{\mathcal{D} \mathcal{S}}(pp_{\mathcal{D} \mathcal{S}})$, $(pk_{tx},sk_{tx})\leftarrow KGen_{\mathcal{C} \mathcal{H}}(pp_{\mathcal{C} \mathcal{H}})$;\\
    $\mathcal{Q}_{sr},\mathcal{Q}_{cr},\mathcal{Q}_{su},\mathcal{Q}_{cu},\mathcal{Q}_{sw},\mathcal{Q}_{cw},\mathcal{Q}_{ac},\mathcal{Q}_{h},\mathcal{Q}_{apt}\leftarrow \emptyset$;\\
    $(\mathcal{P}^{*},r^{*},h^{*},(\sigma _{r}^{*},\sigma _{w}^{*},\sigma _{tx}^{*}),(pk_{w}^{*},pk_{r}^{*},pk_{tx}^{*},pk_{u}^{*}),tx^{*})\leftarrow \mathcal{A}^{\mathcal{O}}(pp_{\mathcal{D} \mathcal{S}},pp_{\mathcal{C} \mathcal{H}},pk_{u},pk_{tx},sk_{tx},pk_{r},pk_{w},pk_{ca})$;\\
    return 1 iff $(Ver_{\mathcal{D} \mathcal{S}}(\sigma _{w}^{*},tx^{*},pk_{w}^{*})=1 \wedge \sum \mathcal{F}(pk_{w})>ts \wedge Ver_{\mathcal{D} \mathcal{S}}(\sigma _{r}^{*},tx^{*},pk_{r}^{*}) \wedge Ver_{\mathcal{C} \mathcal{H}}(pk_{tx}^{*},h^{*},r^{*},tx^{*})=1 \wedge$\\
    $\mathcal{M}(\mathcal{P}^{*},\mathcal{S}^{*})=1)\wedge ((pk_{u}^{*}\notin \mathcal{Q}_{cu} \wedge (h^{*},r^{*},pk_{u}^{*},tx^{*},\mathcal{P}^{*},\sigma _{tx}^{*})\notin \mathcal{Q}_{h})\vee ((pk_{r}^{*}\notin \mathcal{Q}_{cr}\vee AC_{r}^{*}\notin \mathcal{Q}_{ac})\wedge$\\
    $(\sigma _{tx}^{*},pk_{r}^{*},h^{*},r^{*},\cdot,tx^{*})\notin \mathcal{Q}_{apt})) \vee pk_{w}^{*} \notin \mathcal{Q}_{cw}$,\\
    % $Ver_{\mathcal{D} \mathcal{S}}(WG_{w}^{*},tx^{*},PK^{*})=1 \wedge Ver_{\mathcal{D} \mathcal{S}}(\sigma ^{*},tx^{*},pk^{*})=1 \wedge Ver_{\mathcal{C} \mathcal{H}}(pk_{tx}^{*},h^{*},r^{*},tx^{*})=1 \wedge VSel(w^{*},info^{*})=1 \wedge \mathcal{M}(\mathcal{P}^{*},\cdot)=1 \vee ((pk_{u}^{*}\notin \mathcal{Q}_{cu} \wedge pk^{*}=pk_{u}^{*} \wedge (pk^{*},\mathcal{P}^{*},r^{*},h^{*},tx^{*},\sigma ^{*},WG_{w}^{*})\notin \mathcal{Q}_{h})\vee (pk^{*}\notin \mathcal{Q}_{cr} \wedge PK^{*}\notin \mathcal{Q}_{cw} \wedge )\wedge (r^{*},h^{*},\sigma ^{*},pk^{*},tx^{*})\notin \mathcal{Q}_{apt} \wedge (w^{*},info^{*})\notin Sel(\cdot))$,\\
    else return 0.\\
    \hline
    $\mathcal{O}_{1}(pk_{u},tx,\mathcal{P})$\\
    $(h,r)\leftarrow Hash_{\mathcal{C} \mathcal{H}}(pk_{tx},tx,\mathcal{P})$; $\sigma _{tx} \leftarrow Sig_{\mathcal{D} \mathcal{S}}(tx,\mathcal{P},sk_{u})$; $\mathcal{Q}_{h}\leftarrow \mathcal{Q}_{h}\cup (h,r,pk_{u},tx,\mathcal{P},\sigma _{tx})$; return $(h,r,\sigma _{tx})$.\\
    \hline
    $\mathcal{O}_{2}(\sigma _{tx},pk_{r},h,r,tx,tx')$\\
    $r'\leftarrow Apt_{\mathcal{C} \mathcal{H}}(sk_{tx},h,r,tx,tx'), \sigma _{r} \leftarrow Sig_{\mathcal{D} \mathcal{S}}(tx',AC_{r},sk_{r})$; $\mathcal{Q}_{apt}\leftarrow \mathcal{Q}_{apt}\cup (pk_{r},h,r,tx,tx',r',\sigma _{r})$; return $(r',\sigma _{r})$.\\
    \hline
    $\mathcal{O}_{3}(i)$\\
    $(pk_{r},sk_{r})\leftarrow KGen_{\mathcal{D} \mathcal{S}}(pp_{\mathcal{D} \mathcal{S}})$; $\mathcal{Q}_{sr}\leftarrow \mathcal{Q}_{sr}\cup (i,pk_{r},sk_{r})$; return $pk_{r}$.\\
    \hline
    $\mathcal{O}_{4}(pk_{r})$\\
    $\mathcal{Q}_{cr}\leftarrow \mathcal{Q}_{cr}\cup (pk_{r})$; return $sk_{r}$.\\
    \hline
    $\mathcal{O}_{5}(i)$\\
    $(pk_{u},sk_{u})\leftarrow KGen_{\mathcal{D} \mathcal{S}}(pp_{\mathcal{D} \mathcal{S}})$; $\mathcal{Q}_{su}\leftarrow \mathcal{Q}_{su}\cup (i,pk_{u},sk_{u})$; return $pk_{u}$.\\
    \hline
    $\mathcal{O}_{6}(pk_{u})$\\
    $\mathcal{Q}_{cu}\leftarrow \mathcal{Q}_{cu}\cup (pk_{u})$; return $sk_{u}$.\\
    \hline
    $\mathcal{O}_{7}(pk_{r},\mathcal{S})$\\
    $AC_{r} \leftarrow Sig_{\mathcal{D} \mathcal{S}}((pk_{r},\mathcal{S}),sk_{ca})$; $\mathcal{Q}_{ac}\leftarrow \mathcal{Q}_{ac}\cup (pk_{r},\mathcal{S})$; return $AC_{r}$.\\
    \hline
    $\mathcal{O}_{8}(i)$\\
    $(pk_{w},sk_{w})\leftarrow KGen_{\mathcal{D} \mathcal{S}}(pp_{\mathcal{D} \mathcal{S}})$; $\mathcal{Q}_{sw}\leftarrow \mathcal{Q}_{sw}\cup (i,pk_{w},sk_{w})$; return $pk_{w}$.\\
    \hline
    $\mathcal{O}_{9}(pk_{w})$\\
    $\mathcal{Q}_{cw}\leftarrow \mathcal{Q}_{cw}\cup (pk_{w})$; return $sk_{w}$.\\
    \hline
  \end{tabular}
  \setlength{\abovecaptionskip}{0.8cm}
  \setlength{\belowcaptionskip}{0cm}
  \caption*{Figure 2: EUF-CMA security of EFRB Scheme}
  \vspace{-0.1in}
\end{table*}

\section{Experiment and Analysis}
\label{s5}
In this section, we will show the experimental results of the \emph{EFRB} scheme, as well as the analysis of the security and performance of the scheme.

\renewcommand{\thefigure}{3}
\begin{figure*}[t]
	\centering
	\vspace{-0.15in}
	\begin{minipage}{1\linewidth}
		\subfloat[Initialization time]{
			\label{fig:1}
			\includegraphics[width=0.47\linewidth,height=6cm]{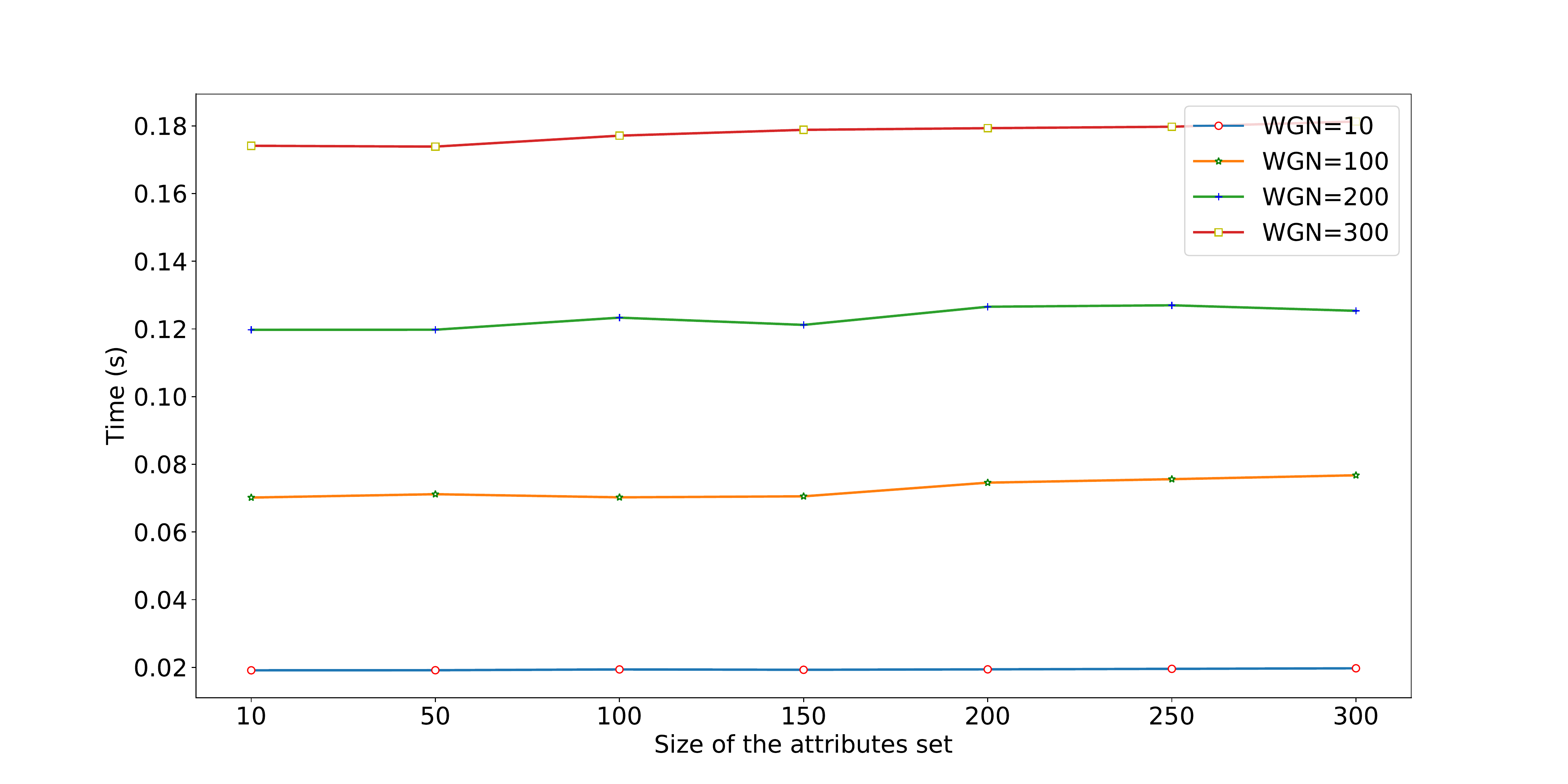}	
		}\noindent
		\subfloat[Time to generate a redactable transaction]{
			\label{fig:2}
			\includegraphics[width=0.47\linewidth,height=6cm]{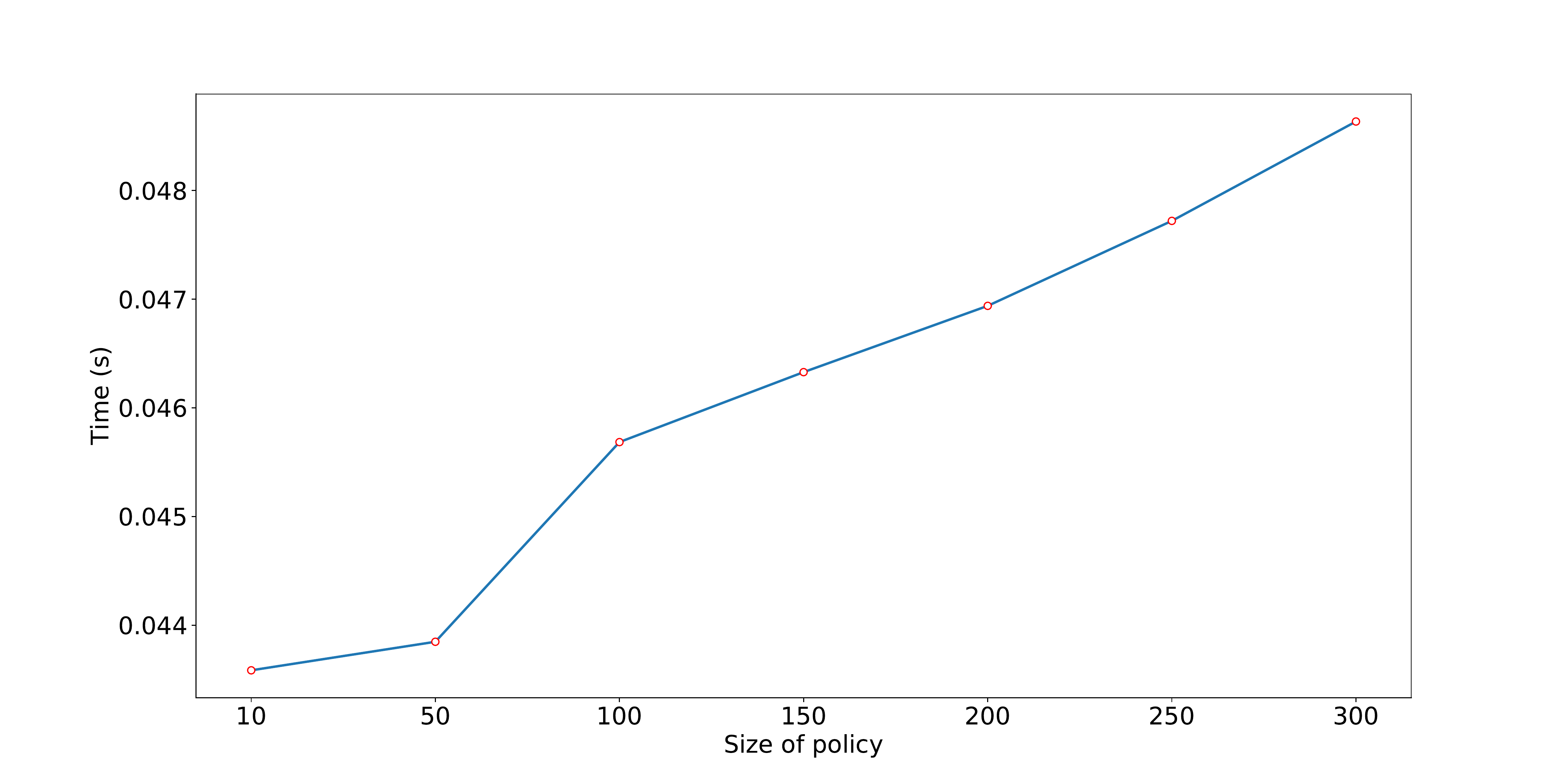}
		}
	\end{minipage}
	\vskip -0.3cm
	\begin{minipage}{1\linewidth }
		\subfloat[Time to generate a redacting request]{
			\label{fig:3}
			\includegraphics[width=0.47\linewidth,height=6cm]{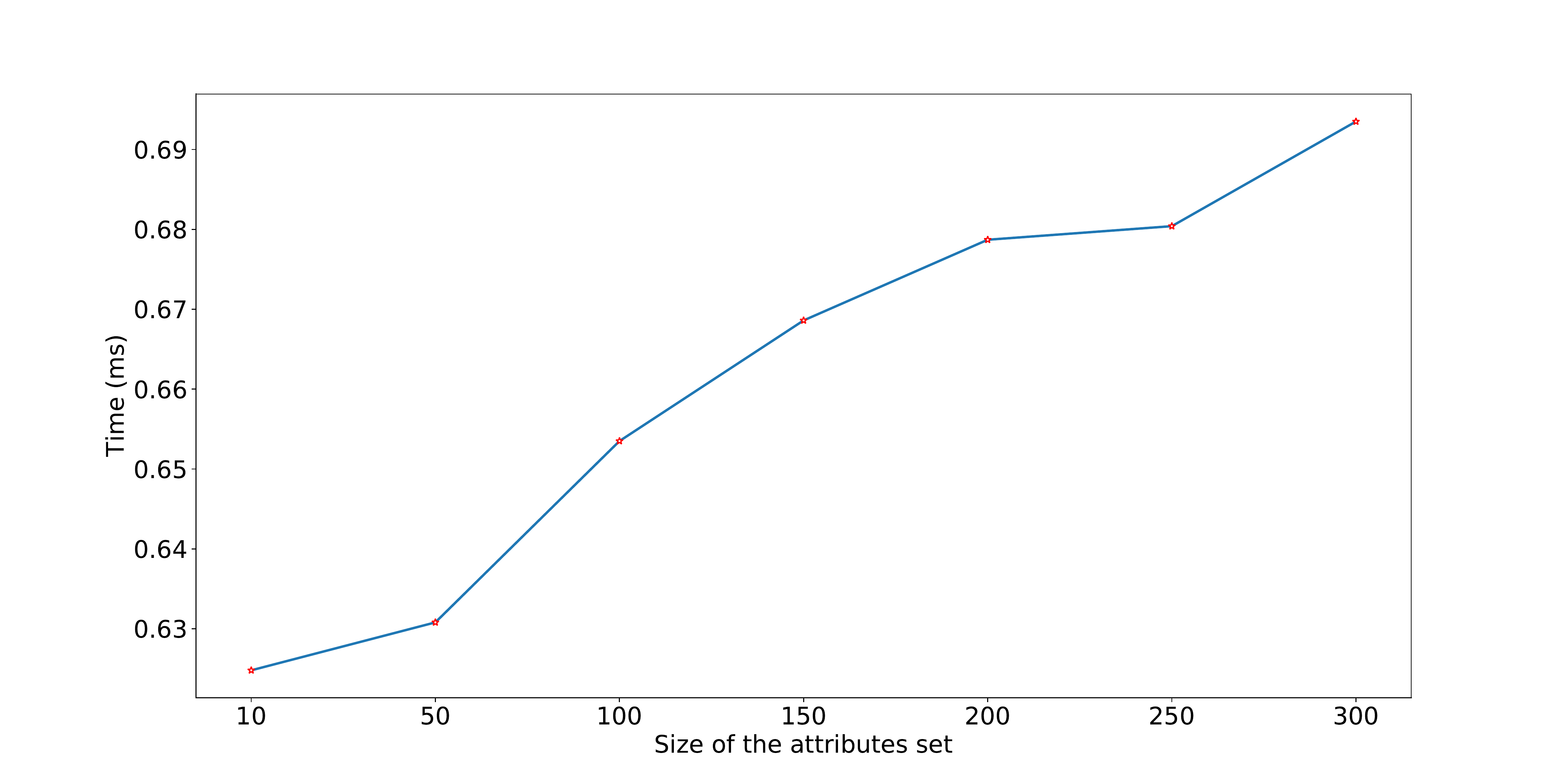}
		}\noindent
		\subfloat[Time to verify a redacted transaction]{
			\label{fig:4}
			\includegraphics[width=0.47\linewidth,height=6cm]{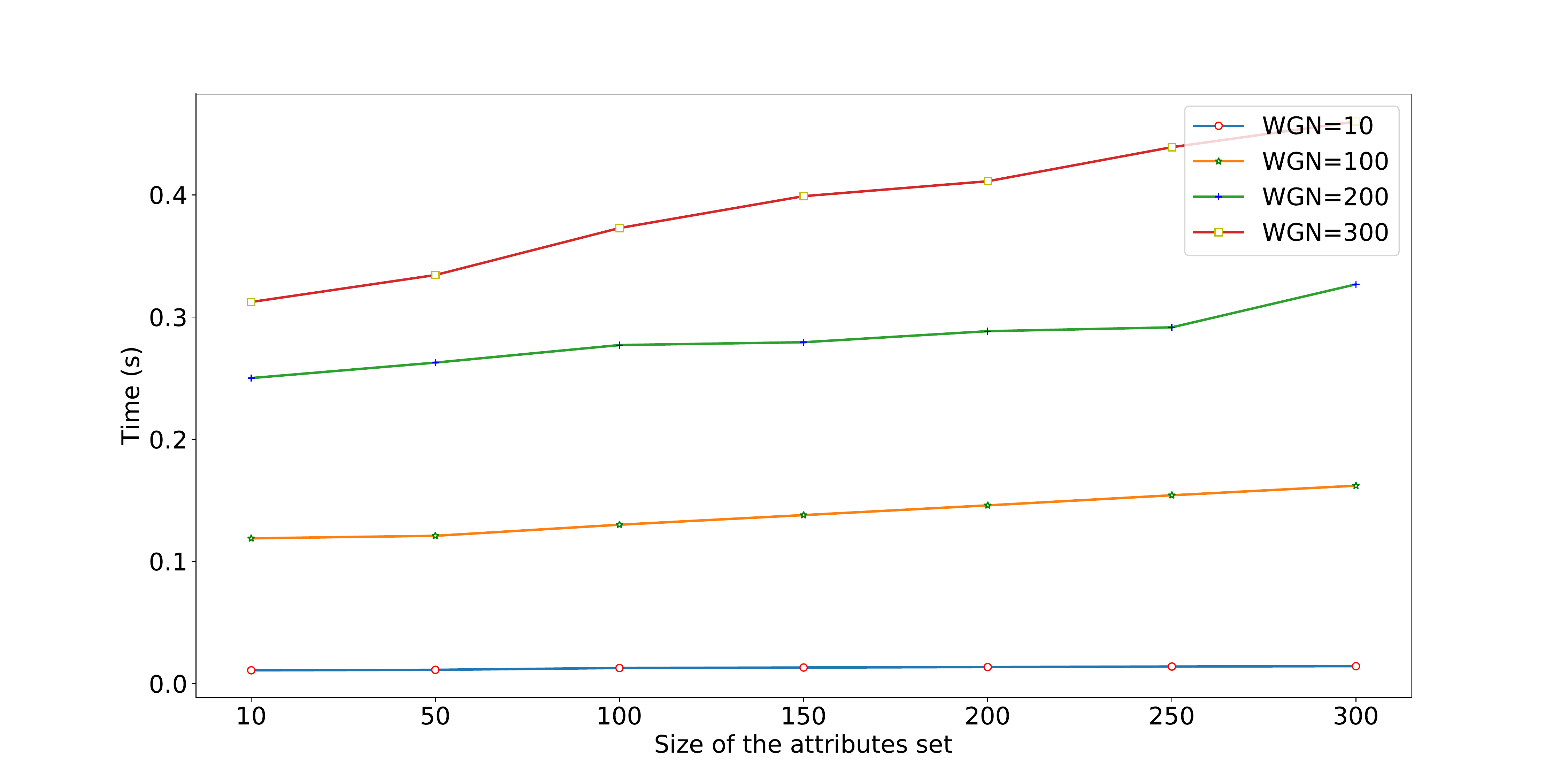}
		}
	\end{minipage}
	\vspace{0.3cm}
	\caption{Experiment of EFRB scheme}
	\vspace{-0.2in}
	\label{fig:1234}
\end{figure*}

\subsection{Security Analysis}
Our scheme mainly uses the chameleon hash function and digital signature, and the trapdoor of the chameleon hash function is public, which means that all users can use it to redact transaction, which can avoid complex trapdoor management problems. In order to control the range of redacting rights, the transaction owner can set a policy bound to the transaction, which specifies which attributes users can become legal redactors of the transaction.
If the owner wants to change the policy, he can also update it through redacting operation. Considering that transaction owners may only set themselves as redactors, or redactors may abuse their redacting rights to illegally tamper with data, we introduce a witness group that acts as a supervisor, so that every redacted transaction needs to be reviewed and signed by the witness group.
The way to achieve the above is to use digital signature: the transaction policy is set in such a way that the transaction owner uses signature to bind the transaction and policy, the authorization of user attributes is that the CA uses signature to bind the attribute and user identity, the redactor needs to sign the transaction data after modifying it, the signature of the witness group is the result of the majority of witnesses using signature as voting.

In addition, the selection method of witness group is closely related to the type of blockchain system, but in essence, it is selected according to the proportion of wealth held by users in the system, such as the stakes in POS and computing power in POW. Like the idea of the underlying consensus mechanism of the blockchain, this is based on the honest majority assumptions, the collision-resistant property of the hash function,
and the fact that users with more wealth are more likely to maintain the system, otherwise they will lose a lot once the system has security problems. Moreover, \cite{gilad2017algorand} and \cite{li2021escaping} prove that it is safe to use similar committee selection methods.

To sum up, if the digital signature has EUF-CMA security and the hash function has collision-resistant property, then the \emph{EFRB} also has EUF-CMA security, that is, $Pr[Exp^{EUF-CMA}_{EFRB,\mathcal{A}}(1^{\kappa})]=Pr[Exp^{EUF-CMA}_{\mathcal{D} \mathcal{S},\mathcal{A}}(1^{\kappa})]=1$ is negligible. In other words, if the underlying digital signature and hash function are secure, our scheme is also secure.

In addition, in order to prevent the witness group from voting lazily, we have added the incentive and punishment mechanism of joint and several liability to let the whole network users supervise the witness group. If some witnesses behave badly during their tenure and are not found by other witnesses, all witnesses in the group will be deducted from the deposit and rewarded to the discoverer.
If the witness group does not behave badly, they will receive the processing fee in the redacting request.

\subsection{Experimental Evaluation}
We implemented the proposed \emph{EFRB} scheme and carried out a series of experiments. The results are shown in Figure \ref{fig:1234}. Specifically, we used Python to construct a simplified blockchain system based on PoW consensus, which simulates some basic blockchain functions. The digital signature we use is ECDSA \cite{johnson2001elliptic} over secp256k1. Our experimental environment is Ubuntu 16.04 (64bits) system, and the configuration is 3.20 GHz AMD Ryzen 7 5800H CPU and 16.0 GB memory.

Firstly, we evaluated the average time required for the scheme initialization, and used \emph{WGN} to represent the size of witness group, as shown in Figure \ref{fig:1}, which mainly includes allocating key pairs to CA, transaction owner, redactor and witness groups, and CA generating attribute certificate for a redactor. It can be seen from the figure that, on the one hand, the initialization time mainly increases with the increase of the number of witnesses, but even if the number of witnesses reaches 300, the initialization time only needs less than 0.2 seconds.
On the other hand, the change in the number of attributes has little impact on the runtime. This is because the attribute set used in the \emph{EFRB} scheme is a string rather than a complex tree structure, which can improve the efficiency of the signature.

Secondly, we evaluated the average time to generate a redactable transaction, as shown in Figure \ref{fig:2}. It can be seen from the figure that the larger the policy set by the transaction owner, the more time it takes to generate a redactable transaction. However, because the policy is also a string, even if a 300 size policy is set, the time required is not more than 0.05 seconds.

Thirdly, we evaluated the average time required to generate a redacting request, as shown in Figure \ref{fig:3}. Because the redactor needs to attach the attribute certificate to the transaction and sign it when generating the redacting request, the more attributes the redactor has, the more time it takes.

Finally, we evaluated the average time for users to verify a redactable transaction, as shown in Figure \ref{fig:4}. Verifying a redactable transaction includes verifying whether the redactor's attributes meet the policy, and whether the signatures of the redactor and the witness group are legal. Therefore, it is the number of witnesses that mainly affects the verification time, followed by the size of the redactor's attribute set.
It can be seen from the figure that the larger the witness group, the more obvious the impact of the size of the attribute set on the verification time. This is because the witness needs to sign the redactable transaction when voting for the redacting request, which includes the redactor's attribute set. Therefore, as the number of witnesses increases, the size of the attribute set will have a greater impact on the verification time.

To sum up, in the \emph{EFRB} scheme, the main factor affecting the efficiency is the size of the witness group. Relatively speaking, the impact of the size of the attribute set on the efficiency can be ignored. Therefore, as long as the size of the witness group is selected reasonably, our scheme can achieve the efficient and fine-grained redacting function of the blockchain with accountability and updatable policies.

\section{Conclusion}
\label{s6}
In this paper, we propose an efficient and fine-grained redactable blockchain scheme with accountability and updatable policies. In our scheme, the transaction owner can set an updatable transaction policy, which enables users whose attributes meet the policy to become the legitimate redactor of the transaction.
In order to prevent redactors from abusing redacting right, we introduce the concept of witness group. The redacted transaction is legal only when the redactor and the witness group have signed the transaction. We first introduce the concept of witness group, and then give the details of the scheme.
Finally, after a series of experiments and analysis, we prove that the scheme is feasible and efficient.

\bibliography{mybibfile}

\end{document}